\documentclass[prd,12pt,nofootinbib,tightenlines]{revtex4}

\usepackage{graphicx}
  \usepackage{amssymb}
   \usepackage{amsmath}
    \usepackage{amsfonts}
     \usepackage{epsfig}
      \usepackage{bm}

\begin{document}

\title{Near-threshold boson pair production in the model\\
of smeared-mass unstable particles}

\author{V.~I.~Kuksa}
\email{kuksa@list.ru} \affiliation{Institute of Physics, Southern
Federal University, Rostov-on-Don, Russia}

\author{R.~S.~Pasechnik}
\email{roman.pasechnik@fysast.uu.se} \affiliation{High Energy
Physics, Department of Physics and Astronomy, Uppsala University,
Box 535, SE-75121 Uppsala, Sweden}

\begin{abstract}
Near-threshold production of boson pairs is considered within the
framework of the model of unstable particles with smeared mass. We
describe the principal aspects of the model and consider the
strategy of calculations including the radiative corrections. The
results of calculations are in good agreement with LEP II data and
Monte-Carlo simulations. Suggested approach significantly simplifies
calculations with respect to the standard perturbative one.
\end{abstract}

%\pacs{11.30.Pb}

\maketitle

%---------------------
\section{Introduction}
%---------------------

Near-threshold production of the unstable particles (UPs) is the
most suitable process to observe the finite-width effects (FWEs).
These effects are closely connected with the instability which
depends on the width and energy of the products $E-E_{thr}$ with
$E_{thr}$ being the threshold energy. It equals to the sum of masses
of the final state particles. The measurements of boson-pair
production at the threshold (e.g. LEP II experiments) have provided
us with an important information about the masses of bosons and
non-abelian triple gauge-boson couplings. To extract the exact
information from the boson-pair production we have to calculate the
radiative corrections (RCs), which give a noticeable contribution to
the corresponding cross-section. Ideally, one would like to have the
full RCs to the processes $e^+e^-\to ZZ,\,ZH,\,W^+W^-\to \sum_f 4f$.
In practice, this problem is very complicated and can not be
considered analytically.

For discussion of the LEP II situation and strategy it is useful to
distinguish three levels of sophistication in description of the
boson-pair production \cite{1,2}:

\begin{enumerate}
\item On-shell boson-pair production, $e^+e^-\to ZZ, ZH, W^+W^-$,
with subsequent on-shell boson decays. All $O(\alpha)$ RCs to these
processes are well-known.

\item Off-shell production of boson pairs which then decay into four
fermions. Full set of the RCs is very bulky for the analytical
observation and analysis.

\item Total process $e^+e^-\to 4f$ with an account of the complete set
of the $O(\alpha)$ corrections. This problem leads to the additional
diagrams with the same final states, and complete electroweak (EW)
$O(\alpha)$ corrections are described by many thousands diagrams.
\end{enumerate}

On-shell $W$-pair production was considered in Refs.~\cite{1,2,3},
where the cross-section of the process $e^+e^-\to W^+W^-$ was given.
At the tree level, this process is described by two $s$-channel
diagrams with $Z,\gamma$-exchange and one $t$-channel diagram with
$\nu$-exchange. The complete $O(\alpha)$ radiative corrections,
comprising the virtual one-loop corrections and real-photon
bremsstrahlung, were calculated and represented in Refs.~\cite{4} --
\cite{11}. The description of the on-shell $W$-pair production and
their subsequent decays with an account of RCs was fulfilled in
Refs.~\cite{12} -- \cite{18}. Off-shell production of $W$-pairs,
which then decay into four fermions, was considered in
Ref.~\cite{19}.

In description of the $W$- and $Z$-pairs production we should take
into consideration the fact that the gauge bosons are not stable
particles and the real process is not $e^+e^-\to W^+W^-,ZZ$
\cite{2}. This is only an approximation with a level of goodness,
which may depend on several factors, while the real process is
$e^+e^-\to W^+W^-,ZZ\to 4f$. There are many papers devoted to
comprehensive analysis and description of all possible processes
with the four-fermion final states. Because of a large number of
diagrams, describing these processes, the classification scheme was
applied in Refs.~\cite{20} -- \cite{23}. The possible processes are
divided into three classes: charge current (CC), neutral current
(NC) and mixed current (MIX). Born processes $e^+e^-\to W^+W^-,ZZ$
are designated as CC03 and NC02, which correspond to three charge
current and two neutral current diagrams. According to this
classification the off-shell $W$-pair production with consequent $W$
decay can be described in the framework of the Double-Pole
Approximation (DPA) \cite{23} -- \cite{26}. The DPA selects only
diagrams with two nearly resonant $W$ bosons and the number of
graphs is considerably reduced \cite{23}.

Complete description of the total set of $4f$-production processes
including radiative corrections is not analytically available due to
a huge number of diagrams and presence of non-factorizable
corrections. But the complete EW $O(\alpha)$ corrections have been
calculated for some exclusive processes, for instance, for the
processes
$e^+e^-\to\nu_{\tau}\tau^+\mu^-\bar{\nu}_{\mu},\,u\bar{d}\mu^-\bar{\nu}_{\mu}$,
and $u\bar{d}s\bar{c}$ \cite{26a,26b}. Because of complexity of the
problem, some approximation schemes are practically applied, namely,
Semi-Analytical Approximation (SAA) \cite{2,26ab}, improved Born
approximation \cite{26bb}, an asymptotic expansion of the
cross-section in powers of the coupling constant \cite{26bbb},
fermion-loop scheme, etc. (see, Introduction in
Refs.~\cite{26a,26b}). There are many computer tools of
calculations, for instance, Monte-Carlo (MC) simulations, such as
RacoonWW \cite{26b,26c,26d} and YFSWW \cite{26e,26f,28}. All above
mentioned methods are based on the traditional quantum field theory
of unstable particles \cite{2}. At the same time, there are some
alternative approaches for description of the UPs such as the
effective theory of UP \cite{29} -- \cite{31}, modified perturbation
theory \cite{31a}, and the model of UPs with smeared mass
\cite{32,32a,32b}.

Now, we consider the effects of finite (large) width of the bosons
$Z$ and $W$, which occur in the vicinity of the threshold. Similar
approach was used for the study of $e^+e^-\to W^+W^-$ in
Ref.~\cite{Ginzburg}. The main feature of the FWEs is the
``smearing'' (fuzzing) of the threshold. In the standard treatment,
this effect is described by taking into account all virtual states
of UP, i.e. its off-shellness. So, the cross-section
$\sigma(e^+e^-\to ZZ)$ is defined as the cross-section of exclusive
four-fermion production $\sigma(e^+e^-\to 4f)$ in DPA. Analogous
definitions can be applied also in the case of another boson-pair
production processes ($W^+W^-,\,ZH,\,Z\gamma$). Such a description
is usually realized with the help of the dressed propagators of UPs.

In this paper, we describe FWEs in the near-threshold boson-pair
production within the framework of the model of UPs with smeared
masses \cite{32a,32b}. The conception of the mass smearing as the
main element of the model is tested by comparison of its predictions
with experimental data on the corresponding cross-sections. In the
second section, we consider the formulation of the mass-smearing
conception and give a short description of the model. In Section 3,
we present the formalism of the model which is used for the
description of the processes with UP in the initial or final state.
Calculation strategy and results are considered in the fourth and
fifth sections.

The main conclusion of our work is the statement that the
mass-smearing conception is in the good agreement with the
experimental data on the near-threshold boson-pair production.
Moreover, this approach leads to a simple and transparent formalism
for description of the processes with participation of unstable
particles.

\section{Smeared-mass unstable particles model}

The model is based on the time-energy uncertainty relation (UR).
Despite of the formal universality, various URs have different
physical nature. This issue has been discussed all the time starting
from Heisenberg formulation of the uncertainty principle (see, for
instance, Refs.~\cite{33} -- \cite{33c} and references therein).
Here, we shortly consider this problem in close analogy with
Ref.~\cite{33c}.

Formally, all URs are based on the Cauchy-Schwarz inequality:
\begin{equation}\label{2.1}
\Delta f \cdot \Delta g \eqslantgtr
\frac{1}{2}\vert\langle\Psi|[\hat{f},\hat{g}]|\Psi\rangle\vert\,,
\end{equation}
where $\hat{f}$ and $\hat{g}$ are the Hermitian operators of some
physical quantities $f$ and $g$, $\Delta f$ and $\Delta g$ are the
standard deviations, and $|\Psi\rangle$ is some vector state. For
example, the Heisenberg UR for momentum and coordinate follows from
Eq.~(\ref{2.1}) and commutation relation
\begin{equation}\label{2.2}
[\hat{p},\hat{q}]=-i\hbar\,\,\,\longrightarrow \,\,\,\Delta
p\cdot\Delta q \eqslantgtr \frac{1}{2}\hbar\,.
\end{equation}
The time-energy UR has a completely different character since time
$t$ is not an operator but parameter in Quantum Mechanics. This
relation follows from Eq.~(\ref{2.1}) and equation for the
time-dependent operator $\hat{Q}(t)$ in the Heisenberg
representation
\begin{equation}\label{2.3}
i\hbar\frac{d\hat{Q}(t)}{dt}=[\hat{Q}(t),\hat{H}],
\end{equation}
where $\hat{H}$ is Hamiltonian (which does not depend on time). From
(\ref{2.1}) and (\ref{2.3}) it follows the formal relation
\begin{equation}\label{2.4}
\Delta E\cdot\Delta t\eqslantgtr \frac{1}{2}\hbar,\,\,\,\Delta
t=\frac{\Delta Q(t)}{|dQ(t)/dt|}\,,
\end{equation}
where $\Delta t$ is the life-time of an excited state \cite{33,33a}.

The first model of UP, based on the time-energy UR, was suggested in
Ref.~\cite{32}. The time-dependent wave function of UP in the rest
frame was written in terms of its Fourier transform as
\begin{equation}\label{2.5}
\Phi(t)\sim \exp\{iMt-\Gamma|t|/2\}\longrightarrow
\frac{\Gamma}{2\pi}\int\frac{\exp\{-imt\}}{(m-M)^2+\Gamma^2/4}dm,
\end{equation}
where $\Gamma=1/\tau$ is the decay width of UP. The right-hand side
of Eq.~(\ref{2.5}) may be interpreted as a mass distribution with a
spread, $\delta m$, related to the mean life $\delta \tau =1/\Gamma$
by the uncertainty relation:
\begin{equation}\label{2.6}
\delta m\cdot\delta\tau\sim 1,\,\,\,\mbox{or}\,\,\,\delta m\sim
\Gamma\,\,\,(c=\hbar =1).
\end{equation}

Thus, from the time-energy UR (\ref{2.4}) for the unstable quantum
system, it follows the conception of UP mass smearing which is
described by UR (\ref{2.6}). Implicit (non-direct) account of the
time-energy uncertainty relation, or instability, is usually
performed by using the complex pole in $S$-matrix or propagator
which describes UP in an intermediate state. Explicit account of
this relation is realized in the description of UP in the final or
initial state with the help of the mass-smearing effect. From
Eq.~(\ref{2.6}) it follows that this effect is noticeable when UP
has a relatively large width. Previously, it was observed in various
fields of particle physics, in particular, in decay processes with
large-width UP participation \cite{32b}, in the boson-pair
production \cite{34,35}, and in the phenomenon of neutrino
oscillations \cite{36,33c}.

Now, let us consider the main ingredients of the model of
smeared-mass unstable particles \cite{32b}. The field function of
the UP can be considered as a superposition of the standard ones,
i.e.
\begin{equation}\label{2.7}
\Phi_a(x)=\int\Phi_a(x,\mu)\,\omega(\mu)\,d\mu,
\end{equation}
where $\omega(\mu)$ is some weight function, and the spectral
component $\Phi_a(x,\mu)$ has the standard form in the case of fixed
mass $m^2=\mu\,:$
\begin{equation}\label{2.8}
 \Phi_a(x,\mu)=\frac{1}{(2\pi)^{3/2}}\int\Phi_a(k)
 \delta(k^2-\mu)e^{ikx}\,dk\,.
\end{equation}

Using representation (\ref{2.8}) we suppose that for an arbitrary
mass parameter $\mu$ the spectral component of the field
$\Phi_a(x,\mu)$ satisfies the Klein-Gordon equation
\begin{equation}\label{2.9}
(\square -\mu)\Phi_a(x,\mu)=0,\,\,\,k^0=\pm\sqrt{k^2+\mu}.
\end{equation}
In another words, within the framework of the model, UP is on the
smeared mass shell characterized by an arbitrary mass parameter
$\mu=k^2$.

The third element of the model is the commutation relations:
\begin{equation}\label{2.10}
 [\dot{\Phi}^{-}_a(\bar{k},\mu),\,\Phi^{+}_b(\bar{q},\mu')]_{\pm}
 =\delta(\mu-\mu') \delta(\bar{k}-\bar{q})\delta_{ab},
\end{equation}
The presence of additional $\delta(\mu-\mu')$ in Eq.~(\ref{2.10})
means an assumption, namely, creation and annihilation of the
unstable particles with various masses do not interfere. The
expressions (\ref{2.7}) -- (\ref{2.10}) are the main elements of the
model under consideration.

The model Green function has a spectral form. In particular, for the
case of scalar UP it reads
 \begin{equation}\label{2.11}
 D(x)=\int D(x,\mu)\,\rho(\mu)\,d\mu,\,\,\,\rho(\mu)=|\omega(\mu)|^2\,,
 \end{equation}
where $D(x,\mu)$ is defined in a standard way for the fixed
$m^2=\mu$, and $\rho(\mu)$ is the probability density of the mass
parameter $\mu$. From the definition (\ref{2.7}) and commutation
relations (\ref{2.10}) it follows that the amplitude of the process
with UP in a final or initial state takes the form
 \begin{equation}\label{2.12}
 A(k,\mu)=\omega(\mu)A^{st}(k,\mu)\,,
 \end{equation}
where $A^{st}(k,\mu)$ is the amplitude at fixed $\mu$ which is
defined in a standard way.

Determination of the weight function $\omega(\mu)$ or corresponding
probability density $\rho(\mu)=|\omega(\mu)|^2$ can be done with the
help of various methods (see Ref.~\cite{32b} for more details). Here
we consider the definition of $\rho(\mu)$ which leads to the
factorisation property of the amplitude \cite{39}.

We match the model propagator of scalar UP to the standard dressed
one as
\begin{equation}\label{2.13}
 \int\frac{\rho(\mu)d\mu}{k^2-\mu+i\epsilon}\longleftrightarrow
 \frac{1}{k^2-M^2_0-\Pi(k^2)}\,,
\end{equation}
where $\Pi(k^2)$ is the conventional polarisation function. It was
shown in Ref.~\cite{32b} that the correspondence (\ref{2.13}) leads
to the following prescription
\begin{equation}\label{2.14}
 \rho(\mu)=\frac{1}{\pi}\,\frac{\mathrm{Im}\,\Pi(\mu)}{[\mu-M^2(\mu)]^2+[\mathrm{Im}\,\Pi(\mu)]^2}\,,
\end{equation}
where $M^2(\mu)=M^2_0+\mathrm{Re}\,\Pi(\mu)$. The relations between
scalar, vector and spinor Green functions following from equations
of motion together with definition (\ref{2.14}) lead to the
correspondences
\begin{equation}\label{2.15}
 \int\frac{-g_{mn}+k_m k_n/\mu}{k^2-\mu+i\epsilon}\rho(\mu)\,d\mu\; \longleftrightarrow\; \frac{-g_{mn}+k_m
 k_n/k^2}{k^2-M^2(k^2)-i\mathrm{Im}\,\Pi(k^2)}\,.
\end{equation}
and
\begin{equation}\label{2.16}
 \int\frac{\hat{k}+\sqrt{\mu}}{k^2-\mu+i\epsilon}\,\rho(\mu)d\mu\;\longleftrightarrow\;
 \frac{\hat{k}+k}{k^2-M^2(k^2)-ik\Sigma(k^2)}\,,
\end{equation}
Note that in Eq.~(\ref{2.16}) we have done the exchange $\Pi(k^2)\to
k\Sigma(k^2)$.

The correspondences (\ref{2.13}) -- (\ref{2.16}) define some
effective theory of UPs. In this theory, the structure of numerators
in Eqs.~(\ref{2.15}) and (\ref{2.16}) differs from the standard one.
The correspondence between the standard and model expressions in the
cases of vector (in unitary gauge) and spinor UP is given by
transition $m\leftrightarrow k$, where $k=\sqrt{k_ik^i}$:
\begin{align}\label{2.17}
&\eta_{mn}(m)=-g_{mn}+k_mk_n/m^2,\,\,\,\hat{\eta}(m)=\hat{k}+m\,\,\,\mbox{(Standard)};\notag\\
&\eta_{mn}(k)=-g_{mn}+k_mk_n/k^2,\,\,\,\hat{\eta}(k)=\hat{k}+k\,\,\,\mbox{(Model)}.
\end{align}
The unstable particles in initial or final states are described by
the following polarisation matrices
\begin{align}\label{2.18}
&\sum_{a=1}^{3} e^a_m(\bar{k})\dot{e}^a_n(\bar{k})=-g_{mn}+\frac{k_mk_n}{k^2}\,\,\,\,\,\mbox{(vector UP)};\notag\\
&\sum_{a=1}^{2}
u^{a,\mp}_i(\bar{k})\bar{u}^{a,\pm}_k(\bar{k})=\frac{1}{2k^0}(\hat{k}+k)_{ik}\,\,\,\,\,\mbox{(spinor
UP)}.
\end{align}
The coincidence of expressions for numerators of propagators
(\ref{2.17}) and polarisation matrices (\ref{2.18}) leads to the
effect of exact factorisation (see Section 3), while the standard
propagators lead to approximate factorisation. This important
property of the model directly leads to the convolution formula for
the decay rates \cite{37} and universal factorized formula for the
cross-sections \cite{38}. The general factorisation method was
suggested in Ref.~\cite{39} on the basis of results in
Refs.~\cite{37,38}.

\section{Cross-section of the boson-pair production}

The processes with UP in initial or final states are described with
the help of the polarisation matrices (\ref{2.18}) and probability
density (\ref{2.14}). Substituting the relation $\mathrm{Im}
\Pi(m)=m\Gamma(m)$ into Eq.~(\ref{2.14}) with $\mu=m^2$, we come up
with the following definition
\begin{equation}\label{3.1}
 \rho(m)=\frac{1}{\pi}\,\frac{m\Gamma(m)}{[m^2-M^2(m)]^2+[m\Gamma(m)]^2}\,,
\end{equation}
where $M^2(m)=M^2_0+\mathrm{Re}\,\Pi(m)$. The value $\Gamma(m)$ is
defined in a standard way by substitution $M\to m$, where $M$ is the
fixed standard mass of the particle and $m$ is variable mass
parameter. In the case $e^+e^-\to Z\gamma$, when there is one UP in
a final state, the model cross-section at the tree level has a
convolution form:
\begin{equation}\label{3.2}
\sigma^{tr}(s)=\int\sigma^{tr}(s,m_Z)\rho(m_Z)\,dm^2_Z,
\end{equation}
where $\sigma^{tr}(s,m_Z)$ is defined in a standard way, $m_Z$ is
variable mass of $Z$-boson and $\rho(m_Z)$ is defined by
Eq.~(\ref{3.1}).

In the case of the boson-pair production $e^+e^-\to ZZ,\,W^+W^-,\,
ZH$, the model cross-section has a double-convolution form:
\begin{equation}\label{3.3}
\sigma^{tr}(s)=\int\int\sigma^{tr} (s,m_1,
m_2)\rho(m_1)\rho(m_2)\,dm^2_1\,dm^2_2,
\end{equation}
where $m_1$ and $m_2$ are variable masses of bosons. The limits of
integrations in Eqs.~(\ref{3.2}) and (\ref{3.3}) will be given in
the next section.

In the framework of the standard treatment the expressions
(\ref{3.2}) and (\ref{3.3}) can be derived as approximations in
convolution method (CM) and semi-analytical approach (SAA). In the
framework of the model these expressions are direct consequences of
model approach (i.e. of the mass-smearing effect). Moreover, as it
was shown in Ref.~\cite{37}, the convolution formula for the
factorized cross-section can be strictly derived for the processes
with UP in an intermediate state. This result is caused by the
effect of exact factorisation of the total process, for instance,
$e^+e^-\to Z\gamma\to f\bar{f}\gamma$. Such an effect makes it
possible to divide the full process into two stages -- the
scattering and decay of the products. Note that such a separation is
exact in the framework of the model under consideration, while in
the standard treatment it is considered as an approximation
(Narrow-Width Approximation).

Now, we illustrate the effect of the threshold smearing in the
process $e^+e^-\to ZZ$, as an example. In Fig.~1 we present the Born
cross-section $\sigma(e^+e^-\to ZZ)$ in the standard approach with
fixed mass $M_Z$ (dashed line) and in the UP model with smeared mass
(solid line). One can see a transparent effect of the threshold
smearing at $\sqrt{s}\approx M_Z$, which gradually disappears with
the increasing of energy. This effect has close analogy with
``standard smearing of threshold'' which caused by virtual states of
$Z$-bosons in the total process $e^+e^-\to ZZ\to f_i\bar{f}_i
f_k\bar{f}_k$ \cite{21}. We show also LEP2 experimental data on the
cross-section with corresponding error bars. From comparison with
these data it follows that the smearing effect improves theoretical
description, however we need to take into account large radiative
corrections (see the next section).
%-------------------------------------------------------------
\begin{figure}[h!]
\centerline{\epsfig{file=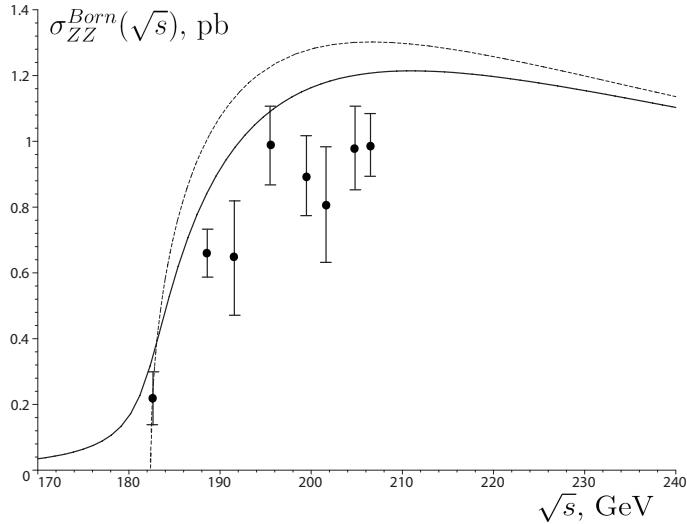,width=9cm}} \caption{Born $ZZ$
cross-section in the Stable Particle Approximation (dashed line) and
in the smeared-mass UP approach (solid line).} \label{fig:Born}
\end{figure}
%-------------------------------------------------------------

\section{Calculation strategy and formalism}

In this section we give the expressions for tree-level
cross-sections and consider the strategy of radiative corrections
(RC) accounting in the framework of the considering model. As it was
shown in Refs.~\cite{32b,39}, the model description of UP is
equivalent to some effective theory of UP, which includes the
self-energy type RCs in all orders of perturbation theory. Moreover,
the UP is the nonperturbative object in the vicinity of the
resonance. So, the traditional program of RCs calculation is not
valid in the framework of the model. We have no well defined set of
the diagrams which is gauge invariant and renormalized. The model of
UPs \cite{32b} is effective and not a gauge one, and we have no any
rigid criteria for definition of such a set. So, we follow the
strategy which is based on the simple phenomenology and was
successfully applied earlier \cite{34,35}. For the preliminary
analysis of the model applicability in description of the boson-pair
production we restrict ourselves by taking into account of the major
part of RCs which is common for all processes under consideration.
This is so-called Initial State Radiation (ISR) correction which
includes hard and soft real and virtual $\gamma$-radiation. It is
needed for compensation of IR divergences.

We do not take into account any corrections to the final states of
UPs because of the effective nature of these states in the framework
of the model. We use the effective couplings $g(M_Z)$ and $\alpha
(M_W)=1/127.9$ in the vertex with the final $Z,\,W,\,H$- and
$\gamma$-states and $\alpha=1/137$ in the calculations of RCs. So,
the principal part of the vertex corrections is effectively included
into the coupling, and the low-energy behavior of the bremsstrahlung
and radiative corrections to the initial states is taken into
consideration.

The set of corrections, caused by the final state interactions in
the $s$-channel diagrams is included into the effective coupling
$\alpha(M_W)$. The principal part of the so-called Coulomb
singularity contributions, which were considered in Refs.~\cite{1},
\cite{26a} and \cite{26bbb}, can be also absorbed by the effective
coupling. The one-loop calculation shows that this correction gives
from $5.7\%$ at the threshold to 1.8\% at 190 GeV \cite{1}, while
the total change of the effective coupling $\alpha(M_W)$ with
respect to $\alpha$ is about $7\%$. In the calculation we explicitly
take into account the $O(\alpha)$ corrections including soft and
hard bremsstrahlung, which are not described by the model and by the
effective coupling. The real and virtual electromagnetic radiation
should enter into the set of these RCs and mutually compensates the
total IR divergences.

The program of RCs calculations, which is similar to above discussed
one, was fulfilled in the series of papers (see, for example,
Ref.~\cite{11} and references therein) for the case of the on-shell
$W$-pair production (the limit of fixed masses $\mu_1=\mu_2=M^2_W$).
The analytical expression for these corrections is represented in
the compact and convenient form in Ref.~\cite{11}. We generalized
this expression to the case of smeared-shell boson-pair production,
that is for arbitrary values of mass parameters $\mu_k$, and applied
it in our calculations. As a result, we get the cross-section
$\sigma_{B_1 B_2}(s;\mu_1,\mu_2)$ in the case of $B_1(\mu_1)$ and
$B_2(\mu_2)$ production including above described corrections in the
following form (see also Ref.~\cite{11})
\begin{equation}\label{4.1}
\sigma_{B_1B_2}(s;\mu_1,\mu_2)= \int_{0}^{k_{max}}
\rho_{\gamma}(k)\sigma^{tr}_{B_1B_2}(s(1-k),\mu_1,\mu_2)\,dk\,,
\end{equation}
where $\rho_{\gamma}(k)$ is the photon radiation spectrum \cite{40}
-- \cite{42}, $k=E_{\gamma}/E_b$ is the photon energy in units of
beam energy and $s(1-k)$ is the effective $s$ available for the
$B$-pair production after the photon has been emitted \cite{11}. In
the case of the on-shell $W$-pair production ($\mu_1=\mu_2=M^2_W$)
the value $k_{max}=1-4M^2_W/s$ is the maximal part of the photon
energy. The generalization of this value to the case $\mu_1\ne\mu_2$
leads to
\begin{equation}\label{4.2}
k_{max}=1-2\frac{\mu_1+\mu_2}{s}+\frac{(\mu_1-\mu_2)^2}{s^2}\equiv
\lambda^2(\mu_1,\mu_2;s).
\end{equation}
The photon distribution function is written in the form \cite{11}
\begin{equation}\label{4.3}
\rho_{\gamma}(k)=\beta
k^{\beta-1}(1+\delta^{v+s}_1+...)+\delta^h_1+...,
\end{equation}
where we keep $O(\alpha)$ corrections only (i.e. $\delta_{n>1}=0$).
The corresponding corrections are given ($v+s=$ virtual+soft, $h=$
hard) in Ref.~\cite{11} by
\begin{align}\label{4.4}
&\beta=\frac{2\alpha}{\pi}(L-1),\,\,\,L=\ln\frac{s}{m^2_e},\,\,\,
\alpha=\frac{1}{137};\notag\\
&\delta^{v+s}_1=\frac{\alpha}{\pi}(\frac{3}{2}L+\frac{\pi^2}{3}-2),\,\,\,
\delta^h_1=\frac{\alpha}{\pi}(1-L)(2-k).
\end{align}
In analogy with Ref.~\cite{11}, we take into account an effective
QCD correction factor in the multiplicative form
$k_{QCD}=1+0.133/\pi$ \cite{43}.

Now, we present the expressions for the cross-sections under
consideration at the tree level. The scattering $e^+e^-\to ZZ$ is
described by two standard $t$-channel diagrams. The model
cross-section differs from the standard one due to various masses of
$Z_1$ and $Z_2$ \cite{34}
\begin{equation}\label{4.5}
 \sigma^{st}(e^+e^-\rightarrow Z(m_1)Z(m_2))=
 \frac{g^4(1+6c^2+c^4)}{2^{10}\pi s\cos^4{\theta_W}}
 \bar{\lambda}(m_1,m_2;\sqrt{s})f(m_1,m_2;\sqrt{s}),
\end{equation}
where $c=1-4\sin^2{\theta_W}$ and $g$ is the weak coupling constant.
The functions $\bar{\lambda}(m_1,m_2;\sqrt{s})$ (normalized K\"allen
function) and $f(m_1,m_2;\sqrt{s})$ are defined by the following
expressions
\begin{equation}\label{4.6}
 \bar{\lambda}(m_1,m_2;\sqrt{s})=\left[1-2\,\frac{m^2_1+m^2_2}{s}+
 \frac{(m^2_1-m^2_2)^2}{s^2}\right]^{1/2}
\end{equation}
and
\begin{equation}\label{4.7}
 f(m_1,m_2;\sqrt{s})=-1+\frac{s^2+(m^2_1+m^2_2)^2}
 {s(s-m^2_1-m^2_2)\bar{\lambda}(m_1,m_2;\sqrt{s})}
 \arctan{\frac{s\bar{\lambda}(m_1,m_2;\sqrt{s})}{s-m^2_1-m^2_2}}.
\end{equation}

The scattering $e^+e^-\to W^+W^-$ is described by one $t$-channel
and two $s$-channel ($\gamma$, which is neglected, and $Z$ in the
intermediate state) standard diagrams. The model cross-section at
the tree level is as follows \cite{35}
\begin{equation}\label{4.8}
\sigma^{tr}_{WW}(s;x_1,x_2)=\frac{\pi\alpha^2}{128 s
\sin^4\theta_{W}}F(s;x_1,x_2)\,,
\end{equation}
where dimensionless function $F(s;x_1,x_2)$ is defined by the
expression
\begin{align}\label{4.9}
F(s;x_1,x_2)&=\frac{16}{3(a^2-b^2)(1-x_Z)^2}\{3(a^2-b^2)(a^2-b^2+2(1+a))(1-x_Z)^2 L(a,b)\notag\\
            &+x_Z\cos(2\theta_W)[3(b^4-2ab^2(2+a)+a^3(4+a))(1-x_Z)L(a,b)\notag\\
            &+2\lambda(a,b)(2b^2-3a^2-10a-1)(b^2(1-2x_Z)-a(1-3x_Z)-x_Z)]\notag\\
            &+\lambda(a,b)[x^2_Z\lambda^2(a,b)\cos(4\theta_W)(2b^2-3a^2-10a-1)+12a^3z^2_Z\notag\\
            &-a^2(3b^2(3x^2_Z-2x_Z+1)-49x^2_Z+30x_Z-15)-2a(b^2(19x^2_Z-10x_Z+5)\notag\\
            &+8x^2_Z)+2b^4(3x^2_Z-2x_Z+1)-2b^2(7x^2_Z-16x_Z+8)-2x^2_Z]\}.
\end{align}
In Eq.~(\ref{4.9}) the dimensionless variables $a,b,x_1,x_2,x_Z$
and the functions $L(a,b)$ and $\lambda(a,b)$ are defined as
follows
\begin{align}\label{4.10}
&L(a,b)=\ln\biggl[\frac{1-a-\lambda(a,b)}{1-a+\lambda(a,b)}\biggr],\,\,\,\lambda(a,b)=\sqrt{1-2a+b^2},\notag\\
&x_{1,2}=\frac{m^2_{1,2}}{s},\,\,\,a=x_1+x_2,\,\,\,b=x_1-x_2,\,\,\,x_Z=\frac{M^2_Z}{s}.
\end{align}

The scattering $e^+e^-\to Z\gamma$ is described by two $t$-channel
standard diagrams. The model cross-section at the tree level
coincides with the standard one
\begin{equation}\label{4.11}
\sigma^{tr}_{\gamma Z}(s,m_Z)=\frac{\alpha g^2}{16\cos^2\theta_W
s}\cdot\frac{1}{1-\mu_Z}\left[(1+c^2_V)(1+\mu^2_Z)[\ln(\frac{s}{m^2_Z}-1]+\frac{4m^2_e}{s}\right],
\end{equation}
where $\mu_Z=m^2_Z/s$.

The scattering $e^+e^-\to ZH$, where $H$ is standard scalar Higgs
boson, is described by one $s$-channel standard diagram with $ZZH$
vertex. This process is the most interesting one from the two points
of view. Besides of its conceptual importance for the Standard Model
verification, this process in the framework of the model has
multiple factorisation structure. From the one hand, due to unstable
$Z$ in the intermediate state, it is described by the universal
factorized formula for the two-particle cross-section \cite{38}. In
the case under consideration, it has a simple factorized form
\begin{equation}\label{4.12}
\sigma^{tr}(e^+e^-\to Z(s)\to Z
H)=\frac{64\pi}{3(1-4m^2_e/s)}\frac{\Gamma^{ee}_Z(s)\Gamma^{ZH}_Z(s)}{(s-M^2_Z)^2+s\Gamma^2_Z(s)}\,,
\end{equation}
where $\Gamma^{ab}_Z(s)$ is the partial width of $Z$-boson, which
has the mass $m^2=s$. Substitution of the expressions for the
$\Gamma^{ee}_Z(s)$ and $\Gamma^{ZH}_Z(s)$ into Eq.~(\ref{4.12})
leads to the final expression for the cross-section in the limit of
zero electron masses:
\begin{equation}\label{4.13}
\sigma^{tr}_{ZH}(s;m_Z,m_H)=\frac{g^4M^2_Z}{108\pi \cos^4
\theta_W}\frac{1-4\sin^2\theta_W+8\sin^4\theta_W}{(s-M^2_Z)^2+s\Gamma^2_Z(s)}
\bar{\lambda}(m^2_Z,m^2_H;s)\Big[1+\frac{(s+m^2_Z-m^2_H)^2}{8sm^2_Z}\Big],
\end{equation}
where $m_Z$ and $m_H$ are variable masses of $Z$-boson and Higgs
boson. The value $\Gamma_Z(s)$ is defined in a standard way with the
change $M^2_Z\to s$.

\section{Results}

In this section we represent the results of the cross-section
calculations in the smeared-shell UP model. In this approach we take
into account the FWEs and the most important RCs (see the previous
section). The model cross-section $\sigma(e^+e^-\to ZZ)$ including
above mentioned corrections is represented in Fig.~2 (the solid
line) together with result of the Monte-Carlo simulation (the dashed
line) and LEP data points \cite{44}. Both results are consistent
with the data within the error bars and coincide with a very high
precision. From this result it follows that the contribution of
non-factorisable corrections in the considered energy range is
negligibly small. So, one can apply our approach to the process
$e^{+}e^{-}\to W^{+}W^{-}$ in this range. The lines start to differ
slightly at energies larger than that of the available data, i.e. at
$\sqrt{s}>200\,\mathrm{GeV}$. Note, however, that the difference
between model and Monte-Carlo curves is an order of differences
between results of various Monte-Carlo calculations.
%-------------------------------------------------------------
\begin{figure}[h!]
\centerline{\epsfig{file=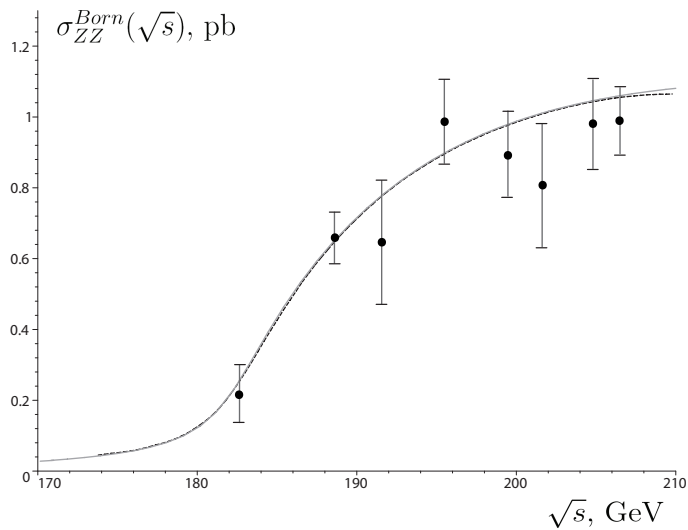,width=9cm}} \caption{Total $ZZ$
cross-section obtained with the Monte-Carlo simulations (dashed
line) and in the model of UP with smeared mass (solid line).}
\label{fig:total}
\end{figure}
%-------------------------------------------------------------

Now, we consider the corrected cross-section of the $W$-pair
production.
%-------------------------------------------------------------
\begin{figure}[h!]
\centerline{\epsfig{file=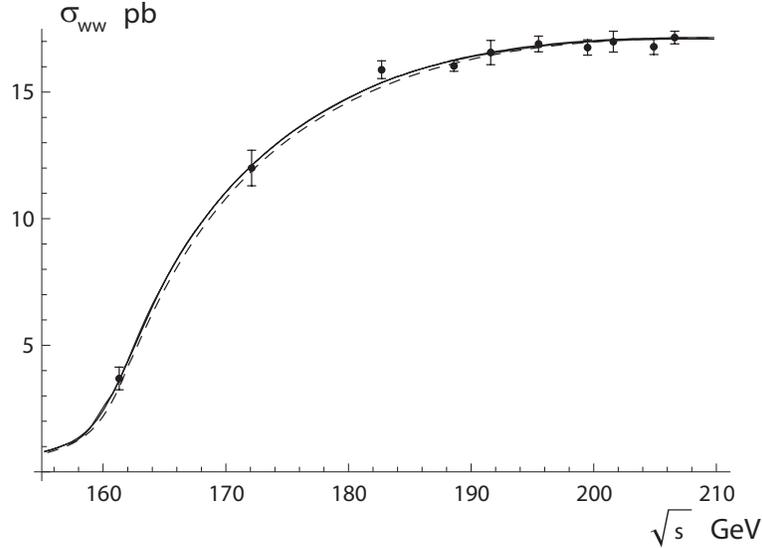,width=10cm}}  \caption{Model
(dashed line) and Monte-Carlo RacconWW and YFSWW (solid lines)
cross-sections of the process $e^+e^-\to W^+W^-$.} \label{fig:FD}
\end{figure}
%-------------------------------------------------------------

The model cross-section $\sigma_{WW}(s)$ was calculated numerically
and represented in Fig.~3 as a function of $s$ by dashed line. The
results of MC simulations, RacconWW \cite{26c,26d} and YFSWW
\cite{26e,26f}, are represented for comparison by two barely
distinguishable solid lines, and the experimental LEP II data
\cite{45} are given with the corresponding error bars. From Fig.~3,
one can see that the model cross-section with RCs is in good
agreement with the experimental data. Moreover, the deviation of the
model from MC curves is significantly less then the experimental
errors ($\lesssim1\%$).

The cross-sections of the process $e^+e^-\to Z\gamma$ are given in
Fig.~4 at the tree level for fixed ($M_Z$, short-dashed curve) and
smeared boson mass (with an account of FWEs, dashed curve). The
corrected cross-section (ISR, effective couplings, etc.) is
represented in this figure by the solid curve. From the results of
calculation it follows that the contribution of the FWEs and RCs is
significant at threshold energy region $\sqrt{s}\gtrsim M_Z$.
Unfortunately, we have no experimental data on the cross-section at
this interesting energy range.
%-------------------------------------------------------------
\begin{figure}[h!]
\centerline{\epsfig{file=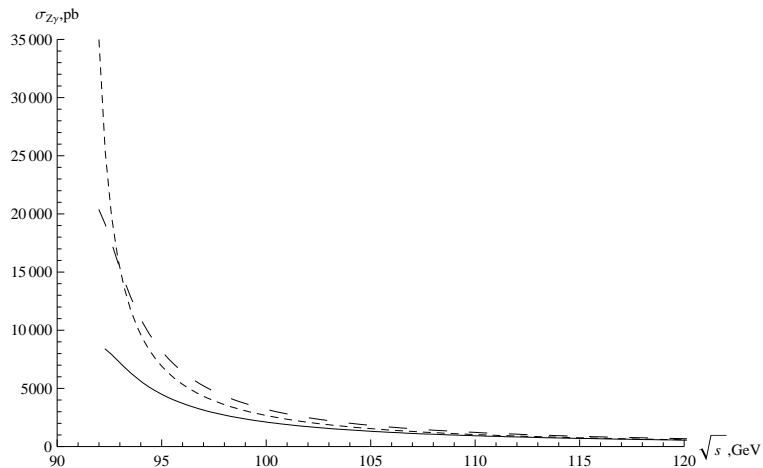,width=10cm}} \caption{Born
$Z\gamma$ cross-section in the Stable Particle Approximation
(short-dashed line) and smeared mass approach (dashed line). Solid
line represents the corrected cross-section.} \label{fig:Born1}
\end{figure}
%-------------------------------------------------------------

The comparison of the model cross-section with the experimental data
was fulfilled for exclusive process $e^+e^-\to \nu\bar{\nu}\gamma$
and $e^+e^-\to q\bar{q}\gamma$ at $160\lesssim\sqrt{s}\gtrsim
200\,\mbox{GeV}$. The model exclusive cross-section of the process
with the final decays $Z\to f\bar{f}$ can be obtained by the
changing $\rho(m)\to \rho(m)Br(Z(m)\to f\bar{f})$ in the formula
(\ref{3.2}). In Fig.~5 we represent the cross-section as a function
of $\sqrt{s}$ within the UP model (solid curve) in comparison with
the experimental data \cite{L3} applying the experimental cuts on
the phase space.
%-------------------------------------------------------------
\begin{figure}[h!]
\centerline{\epsfig{file=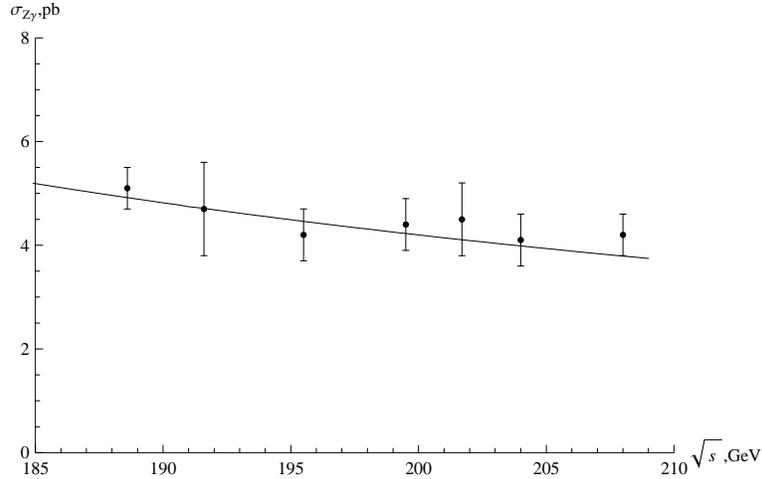,width=10cm}}
\caption{Cross-section of the exclusive process $e^+e^-\to
Z\gamma\to \nu\bar{\nu}\gamma$. Solid line represents corrected
model cross-section.} \label{fig:Born2}
\end{figure}
%-------------------------------------------------------------

One can see that the model description of the process under
consideration is in good agreement with the experimental data. Some
exceeding of the experimental points over the model curve at
high-energy sector can be explained, for instance, by neglecting of
the non-factorisable corrections.
%-------------------------------------------------------------
\begin{figure}[h!]
\centerline{\epsfig{file=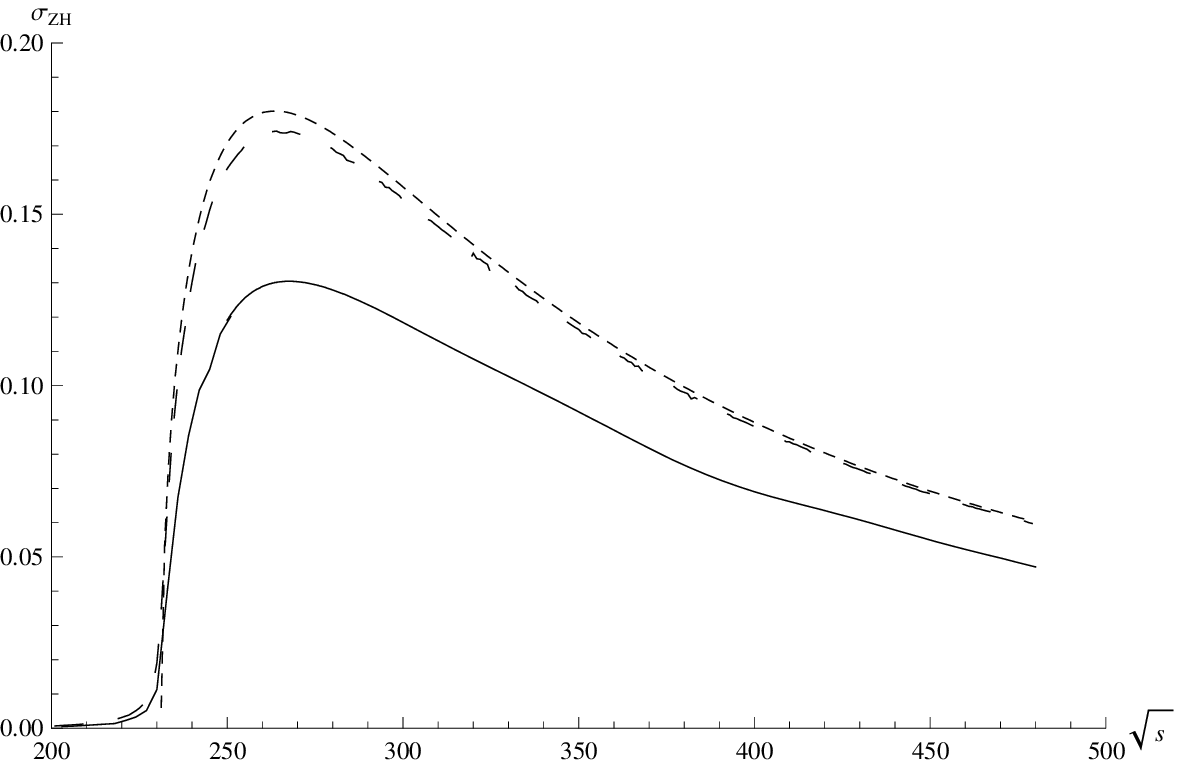,width=10cm}} \caption{Born
$e^+e^-\to Z\to ZH$ cross-section in the Stable Particle
Approximation (short-dashed line) and smeared mass approach (dashed
line). Solid line represents the corrected cross-section.}
\label{fig:ZH}
\end{figure}
%-------------------------------------------------------------

Finally, we consider the cross-section of the process $e^+e^-\to
Z\to ZH$ which is doubly factorisable within the UP model. In the
Fig.~6 we represent the cross-section at the tree level in the case
of fixed boson masses ($M_Z$ and $M_H=140$ GeV, short-dashed line)
and the smeared masses (dashed line). The solid line represents the
corrected model cross-section with an account of the above discussed
RCs. From this figure, one can see the significant role of the
threshold smearing at the threshold energy range. Analogously to
previous case, RCs give quite noticeable contribution, especially in
the peak region at $\sqrt{s}\sim 250 - 300$ GeV.

%-------------------------
\section{conclusion}
%-------------------------

The Finite Width Effects in the processes with participation of the
unstable particles are usually described by the renormalized
propagator, the decay-chain method, the convolution method and by
the effective theory of unstable particles (UPs). In this paper, we
applied the model of UPs with smeared mass for the description of
the boson-pair production. The model describes the process
$e^+e^-\to B_1 B_2$ where bosons are on the smeared mass-shell. This
approach is similar to the standard description of the off-shell
$Z$- and $W$-pair production in the Semi-Analytical Approach. We
have taken into account the soft and hard initial state radiation
and a part of the virtual radiative corrections which are relevant
in the framework of the model.

From our results it follows that the model is applicable to
description of the near-threshold boson-pair production with LEP II
accuracy. We get the total cross-section which is in good accordance
with the experimental data; it coincides with the Monte Carlo
calculations with a high precision. At the same time, the model
provides a compact analytical expression for the cross-section in
terms of convolution of the Born cross section with probability
densities (or mass distributions) of bosons masses. However, we did
not fulfill the detailed analysis of an accounting of the EW
corrections, so this phenomenological formalism can not be directly
applied for the precise description of the boson-pair production at
high energies and for future experiments at ILC. From our results it
follows, that the formalism under consideration can be convenient,
simple and transparent framework for such a description. It is
reasonable to consider the possibility of improvement of the
approach and its applicability in the high precision calculations.

\end{document}